# Revealing unusual bandgap shifts with temperature and bandgap renormalization effect in phase-stabilized metal halide perovskite thin films

Haochen Zhang[1,2,#], Zhixuan Bi[1,#], Zehua Zhai[1,#], Han Gao[3], Yuwei Liu[4], Meiling Jin[4], Meng Ye[5], Xuanzhang Li[1], Haowen Liu[1], Yuegang Zhang[1], Xiang Li[4], Hairen Tan[3,6,*], Yong Xu[1,7,8,*], Luyi Yang[1,2,7,8,*]

[1]State Key Laboratory of Low Dimensional Quantum Physics, Department of Physics, Tsinghua University, Beijing 100084, China

[2]Department of Physics, University of Toronto, Toronto, Ontario M5S 1A7, Canada

[3]National Laboratory of Solid State Microstructures, Collaborative Innovation Center of Advanced Microstructures, Jiangsu Key Laboratory of Artificial Functional Materials, College of Engineering and Applied Sciences, Nanjing University, Nanjing 210093, China

[4]Center for Quantum Physics, School of Physics, Beijing Institute of Technology, Beijing 100081, China

[5]Graduate School of China Academy of Engineering Physics, Beijing 100193, China

[6]Frontiers Science Center for Critical Earth Material Cycling, Nanjing University, Nanjing 210093, China.

[7]Frontier Science Center for Quantum Information, Beijing 100084, China

[8]Collaborative Innovation Center of Quantum Matter, Beijing 100084, China

[#]These authors contributed equally.

[*]e-mail: hairentan@nju.edu.cn; yongxu@mail.tsinghua.edu.cn; luyi-yang@mail.tsinghua.edu.cn

## Abstract

Hybrid organic-inorganic metal halide perovskites are emerging materials in photovoltaics, whose bandgap is one of the most crucial parameters governing their light harvesting performance. Here we present the temperature and photocarrier density dependence of the bandgap in two phase-stabilized perovskite thin films ($MA_{0.3}FA_{0.7}PbI_3$ and $MA_{0.3}FA_{0.7}Pb_{0.5}Sn_{0.5}I_3$) using photoluminescence and absorption spectroscopy. Contrasting bandgap shifts with temperature are observed between the two perovskites. Using X-ray diffraction and in situ high-pressure photoluminescence spectroscopy, we show that thermal expansion plays only a minor role in the large bandgap blueshift, which is attributed to the enhanced structural stability of our samples. Our first-principles calculations further demonstrate the significant impact of thermally induced lattice distortions on the bandgap widening. We propose that the anomalous trends are caused by the competition between static and dynamic distortions. Additionally, both the bandgap renormalization and band filling effects are directly observed for the first time in fluence-dependent photoluminescence measurements and are employed to estimate the exciton effective mass. Our results provide new insights into the basic understanding of thermal and charge-accumulation effects on the band structure of hybrid perovskite thin films.

# 1. Introduction

Over the past few years, compositional engineering has made dramatic breakthroughs in improving the stability and efficiency of thin-film perovskite solar cells[1–3]. Metal halide perovskites, with the formula $ABX_3$ (where A = Cs, methylammonium (MA), or formamidinium (FA); B = Pb or Sn; X = Cl, Br, or I), provide great opportunities to tune their bandgap and structural properties through alloying. While the inorganic octahedral framework $[BX_6]^{4-}$ directly determines the optoelectronic properties of perovskites, the characteristics of the A-site cation (size, motion , spatial distribution, etc.) play an important role in their structural stability by affecting the bond lengths and distortion angles of the inorganic lattice[3,4]. Examples of alloying in perovskites include: (1) partially substituting $FA^+$ (ionic radius 2.53 Å) with the smaller-sized $MA^+$ (2.17 Å), which suppresses the undesired cubic-to-hexagonal transition and stabilizes the photoactive cubic phase at room temperature[5]; (2) partially replacing Pb with Sn reduces the bandgap so that more near-infrared light is absorbed[2]. Both experimental and theoretical studies have pointed out that: (1) for double organic cation ($MA_xFA_{1-x}$) perovskites, the best thermodynamic stability is achieved at $x \approx 0.3$ (Refs.[6,7]); (2) for mixed lead-tin ($Pb_ySn_{1-y}$) perovskites, $y \approx 0.5$ yields the smallest bandgap[8,9].

As light harvesters, one of the most crucial optoelectronic properties of perovskites is the optical bandgap, which is susceptible to temperature and radiation intensity changes. Previous studies have explained the temperature effect on the perovskite's bandgap using the theory developed for conventional semiconductors[10,11], which considers the thermal lattice expansion effect[12–16] and the electron-phonon coupling effect[17–20]. However, compared with conventional semiconductors, perovskites have soft and flexible structures[21–23] and can exhibit lattice distortions such as the octahedral tilting[24–27] and the atom off-centering[28–33]. While both static distortions (the structure at equilibrium) at low temperatures[24,25,34,35] and dynamic lattice vibrations at high temperatures[21–23,36] have been shown to be essential for understanding the lattice dynamics in perovskites, their connection to the thermal evolution of the bandgap remains to be carefully examined.

In addition to the temperature effects, the photo-generated charge accumulation also impacts the bandgap. The state-filling near band edges leads to an increase in the optical bandgap, as described by the Burstein-Moss model, which provides a simple way to estimate the exciton effective mass $\mu$ (Refs.[15,37–39]). The correlation and exchange interactions among photo-injected carriers can lead to a narrowing of the bandgap, a phenomenon known as bandgap renormalization[40,41]. However, only the band-filling effect has been observed in fluence-dependent photoluminescence (PL) measurements so far, causing an overestimate of $\mu$. Therefore, a clear demonstration of both the band-filling and the bandgap renormalization effects is highly desirable.

In this paper, we report on the temperature and photoexcited carrier density dependence of the bandgap in two hybrid perovskite thin films, $MA_{0.3}FA_{0.7}PbI_3$ and $MA_{0.3}FA_{0.7}Pb_{0.5}Sn_{0.5}I_3$ (abbreviated as Pb- and PbSn-perovskite, respectively), using PL and absorption spectroscopy. We chose to study these two compounds for two reasons: (1) the optimization of the organic cation mixing ratio stabilizes the lattice structure, making thermal expansion effects and structural phase transitions unimportant; (2) comparing the conventional pure-lead perovskite with the mixed-lead-tin perovskite can identify critical aspects of the metal cation by examining the difference in their structures. On the application side, our PbSn-perovskite has achieved a record power conversion efficiency when implemented in an all-perovskite tandem solar cell[2,42]. Understanding their bandgap's temperature and photocarrier dependence is essential for enhancing the power conversion efficiency and especially for advancing photovoltaic applications under extreme conditions (e.g., in space).

In our experiments, we observe strong temperature dependence and contrasting behavior of the bandgap between the two perovskites: the bandgap of the PbSn-perovskite shifts almost monotonically by 184 meV from 10 K to 340 K, whereas the bandgap of the Pb-perovskite shifts nonmonotonically with an overall change of only 2 meV in the same temperature range. By performing X-ray diffraction and hydrostatic pressure PL experiments, we find that both samples remain in the cubic phase at low temperatures and the thermal expansion effect only contributes up to ~10% of the bandgap change in our samples. Given their thermally-stable cubic structure, we propose that both static and dynamic lattice distortions are the origin of their unusual bandgap shifts with temperature based on first-principles calculations. In addition, we demonstrate both the bandgap renormalization and the band-filling effects for the first time in fluence-dependent PL measurements, providing a simple and relatively accurate way to determine exciton effective masses. Our study deepens the understanding of the fundamental bandgap and distortion properties in perovskite thin films and provides insights for further engineering of the electronic structure of this emerging class of semiconductors.

## 2. Results and Discussion

### 2.1. Cubic Phase Stabilization

We synthesized high-quality Pb- and PbSn-perovskite thin films based on our previously reported method[42]. Figure 1a presents a schematic of the atom distribution in the mixed FA-MA perovskites. Perovskites commonly undergo several structural phase transitions upon cooling, such as the cubic-to-tetragonal transition at 330 K and the tetragonal-to-orthorhombic transition at 160 K in MAPbI$_3$ (Ref.[5]). Similar transitions also occur in FAPbI$_3$, which suffers from the spontaneous transitions from the α-phase (photoactive) to the δ-phase (non-photoactive) at room temperature and further to the tetragonal phase at 140 K (Ref.[5]). Although our samples have been found to be thermodynamically stable in the cubic phase at room temperature[42], their structure at low temperatures remains unexplored. Previous studies of single-crystal FA$_{0.7}$MA$_{0.3}$PbI$_3$ have identified the cubic $Pm\bar{3}m$ → tetragonal-II $P4/mbm$ → tetragonal-III transition sequence based on the PL shifting curve and the single crystal XRD pattern change with temperature[43,44]. However, the results were inconclusive regarding their low-temperature space group symmetry, and therefore are worth further investigation.

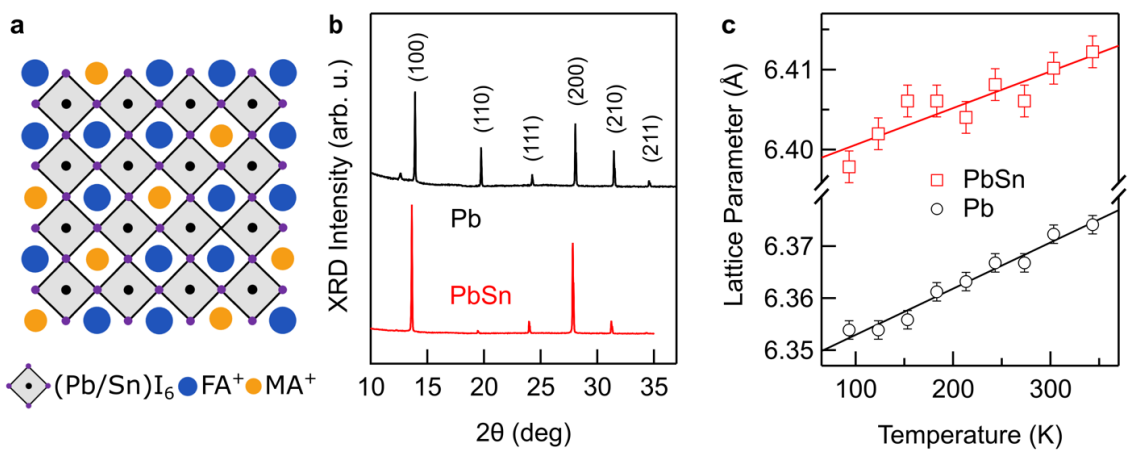

**Figure 1.** Structural properties of the Pb- and PbSn-perovskites. **a** Schematic representation of the atom distribution in mixed-cation perovskite films. **b** Powder X-ray diffraction (PXRD) patterns for the Pb- (black curve) and PbSn- (red curve) perovskites at 90 K. The Miller indices are labelled for each peak. The patterns show the cubic structure for both perovskites. **c** Temperature dependence of lattice parameters in the Pb- (black) and PbSn- (red) perovskites, extracted from PXRD peaks using Bragg's law. The solid lines represent linear fits to the data.

We performed powder X-ray diffraction (PXRD) measurements on the Pb- and PbSn-perovskites from 90 K to 340 K. In contrast to the previous study using single-crystal XRD spectroscopy[44], our PXRD patterns match with the cubic structure over the entire temperature range for both samples (only the 90 K data is shown in Figure 1b for simplicity; the entire temperature-dependent dataset and more details are included in Supporting Note 1). No abrupt change in the diffraction peak intensities or angles is observed, and no additional peak broadening occurs upon cooling. Therefore, we expect long-range cubic symmetry for both our perovskites from 340 K down to 90 K.

The disappearance of phase transitions has also been observed in other cation-alloyed perovskites, such as the mixed methylammonium/dimethylammonium $MA_{0.79}DMA_{0.21}PbBr_3$ hybrid perovskites[45]. Previous research has shown that the A-site cation in the $ABX_3$ perovskite structure directly affects the phase and thermal stabilities in three ways. First, the phase stability can be optimized by the Goldschmidt tolerance factor, defined as $t = (r_A + r_X)/[\sqrt{2}(r_B + r_X)]$, where $r_A$, $r_B$ and $r_X$ are the ionic radii in the A, B, and X sites, respectively. Partially substituting $FA^+$ in $FAPbI_3$ with the smaller-sized $MA^+$ and $Cs^+$ tunes the tolerance factor to the optimized cubic-phase value (0.9 < $t$ < 1) (Refs. [4,5]). Second, since the structural transitions are initiated by the collective tilting of the $[BX_6]^{4-}$ octahedra upon cooling, cation mixing can introduce frustrated electric dipoles that suppress the long-range ordering of the $[BX_6]^{4-}$ octahedra. This results in the stabilization of the cubic phase even at low temperatures[45]. Third, cation mixing may help to relax the crystal strain that builds up during the phase transition[3,4]. Considering that our XRD results are inconsistent with previous single-crystal studies[43,44], it is possible that the potential strain relaxation at crystallite grain boundaries also contributes to the phase stabilization in our thin film samples.

In addition to identifying the structure, the volumetric thermal expansion coefficient ($\alpha_V$) can be inferred from the gradual shift of the PXRD peak angles with increasing temperature. $\alpha_V$ of the Pb- and PbSn-perovskites are 5.6x10$^{-5}$ K$^{-1}$ and 2.9x10$^{-5}$ K$^{-1}$, respectively (Figure 1c). These values are significantly smaller than those of perovskites containing single organic cations[29] (e.g., $\alpha_V$ in the Pb-perovskite is smaller than that in α-$FAPbI_3$ and $MAPbI_3$ by a factor of two) and the mixed-cation perovskite single crystal $MA_{0.5}FA_{0.5}PbI_3$ (Ref.[44]), which again signifies the enhanced structural stability of our samples. However, we have to be aware that $\alpha_V$ of the Pb- and PbSn-perovskites are one order of magnitude higher than those of conventional semiconductors such as Si and GaAs (Ref.[46]), whose covalent bonds are more rigid. The relatively soft and flexible perovskite lattice makes them prone to distorted bond angles while maintaining their cubic structure. The effects of thermal expansion and lattice distortion on bandgap will be discussed in detail later.

## 2.2. Bandgap Shifts with Temperature

To measure the temperature dependence of the bandgap, we record the steady-state PL spectra of the Pb- and PbSn-perovskites from 10 K to 340 K, as shown in Figure 2a,b. The PL peak energy (white circles) shows contrasting temperature dependence between the Pb- and PbSn-perovskites. In the Pb-perovskite, the PL shifts non-monotonically as the temperature increases, with blueshifts below 100 K and above 200 K and redshifts in between, leading to a small overall energy increase of 2 meV in the entire measured temperature range. In the PbSn-perovskite, by contrast, the PL monotonically blueshifts with increasing temperature, and a large overall energy increase of 184 meV is observed in the same temperature range. Additionally, below 70 K, where the bandgap shifts approximately linearly with the temperature, the PbSn-perovskite shows a larger shifting rate of 0.61 meV K$^{-1}$ than that of the Pb-perovskite (0.25 meV K$^{-1}$). From 290 K to 340 K, the shifting rate for the Pb- and PbSn-perovskites is 0.46 meV K$^{-1}$ and 0.72 meV K$^{-1}$, respectively.

The PL energy shift is attributed to the temperature dependence of the bandgap, as confirmed by the absorbance spectra shown in Figure 2c,d. The brown region tracks the absorption onset, and the bandgap values extracted from the Elliot model and the Tauc plot are plotted with black circles (see Supporting Note 3 for details). The bandgap has the same shifting trends as the PL peak energy, but is at a slightly higher energy due to the exciton binding energy[47–51] and the Stokes shift[52], etc. The bandgap blueshift with increasing temperature is an unusual behavior compared with conventional semiconductors, but is ubiquitous in metal halide perovskites. In particular, the large monotonic blueshift of the bandgap seems to be a unique feature in tin-based perovskites[12,53]. In our phase-stabilized Pb-perovskite, a change from blue to red shift in the bandgap is observed between 100 K and 200 K while the structure remains in the cubic phase, indicating that the nonmonotonic bandgap shift cannot be ascribed to phase transitions, as suggested by previous work[13].

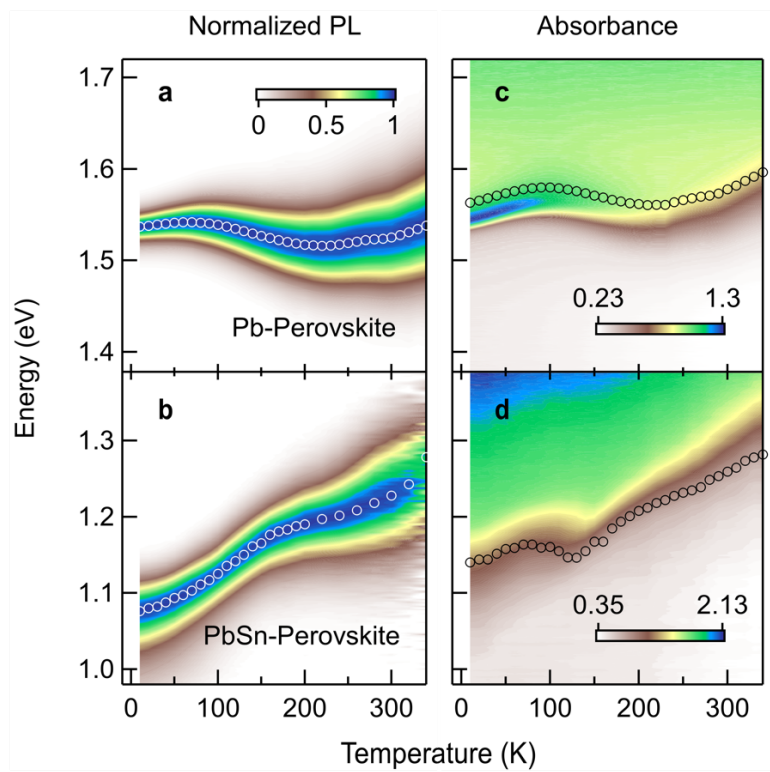

**Figure 2.** Bandgap shift with temperature. Color plots of the normalized PL (**a,b**) and absorbance (**c,d**) spectra, respectively, for the Pb- and PbSn-perovskites from 10 to 340 K. The white open circles in the PL color plots label the PL peak energies and the black circles in the absorbance plots represent the bandgap values extracted from the Elliot model fit (for the Pb-perovskite) and the Tauc plot analysis (for the PbSn-perovskite) of the absorbance spectra (Figure S4, Supporting Information). Error bars are smaller than the markers.

## 2.3. Estimation of Thermal Expansion Effect

Previous analyses of the bandgap shift with temperature in hybrid metal halide perovskites ascribed substantial importance to the thermal expansion effect, both in pure-lead and pure-tin perovskites[12,13,16], due to the large thermal expansion coefficients of the single-organic-cation samples. To make a more accurate estimate of the lattice dilation effect in our samples, we perform hydrostatic pressure PL experiments to measure the pressure coefficient of the bandgap $dE_g/dP$ up to 1.2 GPa at room temperature. As shown in Figure 3, the bandgap shrinks upon increasing pressure in both samples. $dE_g/dP$ is measured to be -41 meV GPa$^{-1}$ and -82 meV GPa$^{-1}$ for the Pb- and PbSn-perovskites, respectively.

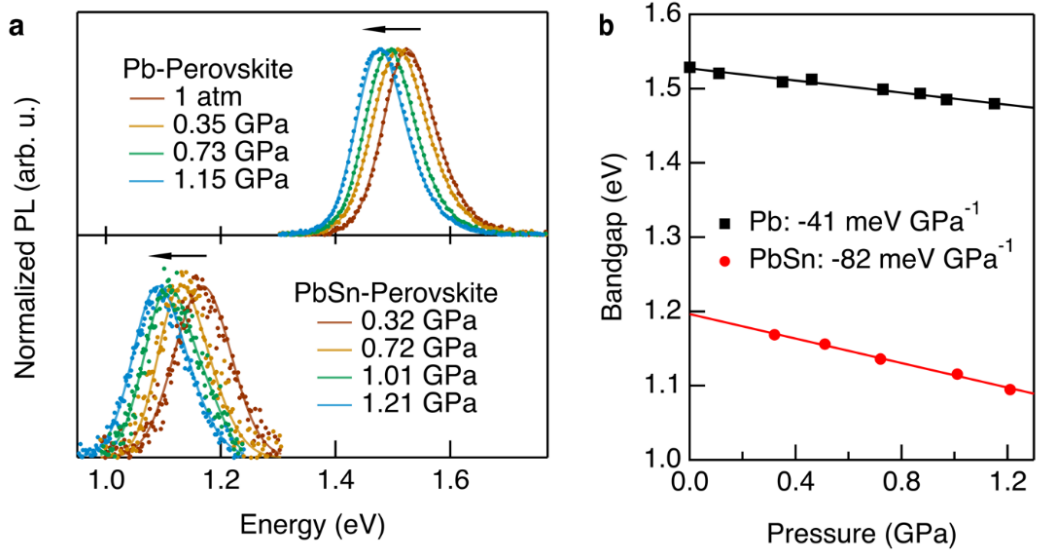

**Figure 3.** Bandgap redshift upon compression at room temperature. **a** PL spectrum with increasing pressures measured in situ in a diamond anvil cell apparatus. The black arrow indicates the evolution of the PL spectra upon compression. **b** Pressure dependence of bandgap deduced from the center of the emissions. Bandgap pressure coefficients are obtained from the slopes of the linear fits and are included in the legend.

The thermal expansion effect on the bandgap is calculated as[54] $(\partial E_g/\partial T)_{TE} = -\alpha_V B_0 (dE_g/dP)$, where $\alpha_V$ is the volumetric thermal expansion coefficient and $B_0$ is the bulk modulus. The pressure coefficients of the bandgap for our samples are comparable to the literature values for FAPbI$_3$ and MAPbI$_3$ perovskites[54,55]. Because in situ synchrotron PXRD experiments under pressure showed similar values of the bulk moduli among all types of iodide perovskites (Table S1, Supporting Information), we use the largest value of $B_0 = 13.6$ GPa found in the literature[56,57] to make an estimate. Then $(\partial E_g/\partial T)_{TE}$ is calculated to be 0.031 meV K$^{-1}$ and 0.033 meV K$^{-1}$ for the Pb- and PbSn-perovskites, respectively, which are significantly smaller than the measured values of 0.25 meV K$^{-1}$ and 0.61 meV K$^{-1}$ at temperatures below 70 K, and 0.46 meV K$^{-1}$ and 0.72 meV K$^{-1}$ at temperatures above 290 K. Therefore, thermal lattice expansion plays a minor role in the bandgap shift of our phase-stabilized perovskites with temperature. Other structural changes are likely responsible for the anomalous bandgap shift. We use numerical methods to analyze these effects in the next section.

## 2.4. Effects of Lattice Distortions on Bandgap

The Pb- and PbSn-perovskites, although stabilized in the cubic phase, possess flexible corner-sharing [PbI$_6$]$^{4-}$ (or [SnI$_6$]$^{4-}$) octahedral frames that are susceptible to distortions. For example, neutron powder diffraction experiments on the tetragonal phase of MAPbI$_3$ have measured the statically distorted I-Pb-I bond angle to be as small as 157° at low temperatures[34,35], with 180° representing the undistorted angle. In addition, strong dynamic distortions at high temperatures have been revealed by molecular dynamics and density functional theory (DFT)[22,36,58], and X-ray and neutron scattering experiments[21,23]. Notably, the dynamic local symmetry breaking caused by highly anharmonic phonons has been revealed in the high-temperature cubic perovskite lattice[21,23,59]. While our PXRD results confirm the preservation of perfect long-range cubic symmetry down to low temperatures, local distortions can still exist. We suspect that the combined effects of static and dynamic lattice distortions with temperature are the most likely origin of the unusual bandgap shifts in our phase-stabilized samples.

To consider appropriate distortion modes in our systems, we first rule out the tilting of the rigid [PbI$_6$]$^{4-}$ (or [SnI$_6$]$^{4-}$) octahedra for the following reasons: (1) static tilting breaks the cubic symmetry, inconsistent with the observed PXRD results; (2) the activation of more dynamic tilting modes is associated with negative thermal lattice expansion and a positive gap pressure coefficient[27], which are opposite in sign to those observed in our Pb- and PbSn-perovskites.

Instead, we examine the effects of the distorted I-Pb-I bond angle with a fixed overall lattice parameter, as depicted in the inset of Figure 4a. This distortion mode (such as the lead atom off-centering effect) has been studied in perovskites before and is sometimes termed "emphanisis" in the literature[28–33]. We present a numerical analysis of the unusual bandgap shifts based on this type of lattice distortion, consisting of two parts: (1) the influence of the lattice distortions on the perovskite band structure, and (2) how (static and dynamic) lattice distortions change with rising temperature.

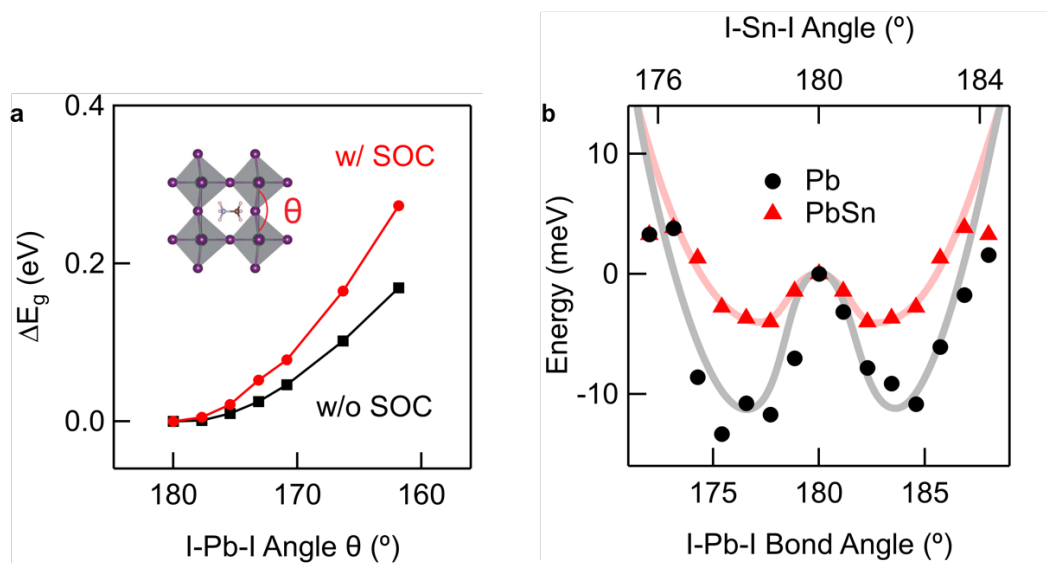

**Figure 4.** Results of first-principles calculations. **a** Bandgap enhancement by lattice distortions represented by the I-Pb-I bond angle $\theta$ (inset, the distortion is applied to the iodine atoms) with (w/) or without (w/o) the SOC effect. **b** Double-well shaped potential of the Pb- and PbSn-perovskites versus the distortion angle. The data for the PbSn-perovskite is extracted from Figure S9 in Supporting Information. The lines in **b** are guides to the eye showing the double well shape of the total energy.

Band structures of the Pb-perovskite calculated using DFT are depicted in Figure S7 (Supporting Information). Because the electronic states near the Fermi level are mainly contributed by the inorganic component, we take the representative composition of MAPbI$_3$ in our calculations for simplicity, while using the cubic lattice structure and lattice parameter from the PXRD measurements (FAPbI$_3$ shows similar results). The conduction band minimum (CBM) corresponds to an anti-bonding state mainly formed by the *p* orbitals of Pb and I, and the valence band maximum (VBM) corresponds to an anti-bonding state mainly formed by the *s* orbital of Pb and the *p* orbital of I.

The strong dependence of the bandgap on the lattice distortion angle is plotted in Figure 4a. The bandgap enhancement reaches hundreds of meV, both with and without spin-orbit coupling (SOC). The lattice distortions denoted in the inset of Figure 4a lead to increased bond lengths, reducing the orbital overlap integrals in the Pb-I bonds. Because antibonding orbitals are repulsively formed with an energy increase, the antibonding CBM and VBM states shift downwards in energy when the overlap reduction occurs with a weaker repulsion. However, the VBM shift is larger than the CBM shift because the Pb *s* orbital of the VBM is more localized than the Pb *p* orbital of the CBM and hence more sensitive to the bond stretching.

In addition to the lattice distortions considered in Figure 4a, we also study another kind of lattice distortions that displaces iodine atoms away from the equilibrium position along the Pb-I bond direction and find similar enhancement of bandgap by lattice distortion as (Figure S6, Supporting Information). Additionally, Figure 4a shows a greater enhancement of the bandgap when SOC is included than when it is not included, which is due to the reduction in the SOC splitting energy with increasing distortion caused by the crystal field effects (see details in Supporting Note 7). Overall, the lattice distortions affect both the orbital overlap and the SOC splitting, leading to a large increase in the bandgap.

To reveal the relationship between lattice distortions and temperature, we calculate the total energy of the system at different distortion angles for both the Pb- and PbSn-perovskites in Figure 4b. The potential energy curves show double-well shapes, nearly symmetric about the undistorted 180° bond angle. The two perovskites have different potential well depths (-11 meV for the Pb-perovskite compared to -4 meV for the PbSn-perovskite) and different distortion angles at the well bottoms (176° and 184° for the Pb-perovskite compared to 178° and 182° for the PbSn-perovskite).

The double-well potential energy curves indicate that the perovskite prefers the distorted angle at the minimum energy at low temperatures. However, as the temperature increases, the lattice structure gradually moves away from the distorted angle and becomes the perfect cubic structure at room temperature. This change in the equilibrium structure reduces the "static" lattice distortion with increasing temperature, thereby shrinking the bandgap. This is supported by a previous study on CsSnI$_3$ (Ref. [59]), which showed that the angles of the double well potential minima shift towards the undistorted angle and the well depth decreases as the thermal vibration is enhanced with increasing temperature.

In addition to "static" deformation, thermally-induced lattice vibrations lead to "dynamic" structural distortions. Previous molecular dynamic simulations of MAPbI$_3$ have shown strong spatial fluctuations of the iodine atoms[60], evidencing significant dynamic distortions in the high-temperature cubic phase. The anharmonic double-well potential has also been suggested to be responsible for the soft phonon modes and the corresponding structural instability at high temperatures[21–23]. Because the lattice vibration amplitude generally grows with temperature, the enhanced dynamic distortions consequently cause an enhanced bandgap. The overall bandgap change with temperature would therefore depend on both the static and dynamic distortion changes.

The PbSn-perovskite likely has a small static but large dynamic lattice distortion change with temperature. Due to its small distortion angle, a small amount of thermal energy can excite the system, resulting in a minor static deformation change with increasing temperature. Additionally, because of the shallow potential well depth (4 meV equivalent to ~ 46 K), the PbSn-perovskite is prone to stronger lattice vibrations and therefore a larger dynamic distortion change. This agrees well with the observation of the much larger bandgap shift rate in the PbSn-perovskite (0.61 meV K$^{-1}$) than the Pb-perovskite (0.25 meV K$^{-1}$) below 70 K (Figure 2). As the temperature increases, the dynamic distortions quickly overcome the static counterpart and dominate, causing the overall monotonic blueshift of the bandgap (Figure 2b,d).

In contrast, the deeper potential well in the Pb-perovskite (11 meV, ~ 126 K) suggests that the system oscillates around one of the distorted angles of the energy minimum at low temperatures and the change of static distortion is very small. Therefore, the bandgap blueshifts due to the enhanced thermal vibrations. In the intermediate temperature range, we speculate that the bandgap redshift between 100 K and 200 K (Figure 2a,c) is due to the substantial reduction of the large static distortion. Note that similar bandgap blue-to-red shift turnover of other lead-based perovskites in the intermediate temperature range has been attributed to the structural phase transitions[13,43], which clearly fails to describe our samples due to their stabilized cubic structures. As the temperature increases further, the local static distortion disappears due to the fast rotation of the A-site cations, and the dynamic distortions regain dominance, leading to an expansion of the bandgap. The moderate overall increase of the bandgap in the Pb-perovskite is due to the compensation between the static and dynamic contributions from 10 K to 340 K.

## 2.5. Bandgap Renormalization and Band-filling Effects

In addition to the temperature dependence, energy shifts of the PL emission are also observed as the excitation density changes. Figure 5a,b shows the shift of PL spectra at 10 K as a function of the pump fluence for the Pb- and PbSn-perovskites (from 3 nJ cm$^{-2}$ to 5 µJ cm$^{-2}$ for the Pb-perovskite, and from 129 nJ cm$^{-2}$ to 15 µJ cm$^{-2}$ for the PbSn-perovskite). The fluence can be converted to photocarrier density $n$ (Supporting Note 8), and the different measurement ranges for the two samples are determined by the detector sensitivity and laser output power (see Experimental Section).

As shown in Figure 5a,c, the Pb-perovskite shows a redshift of the emission under low excitation fluences (up to $n$~$10^{16}$ cm$^{-3}$) followed by a blueshift upon further increasing the pump fluence. In the PbSn-perovskite, however, a monotonic blue shift is observed with increasing photocarrier density ($n$ from $10^{16}$ to $10^{18}$ cm$^{-3}$) (Figure 5b,d). We attribute the PL blueshift to the band filling effect and the redshift to the bandgap renormalization (BGR) effect. The BGR has been observed in perovskites only by transient absorption spectroscopy[61,62] but not with linear optical methods until now.

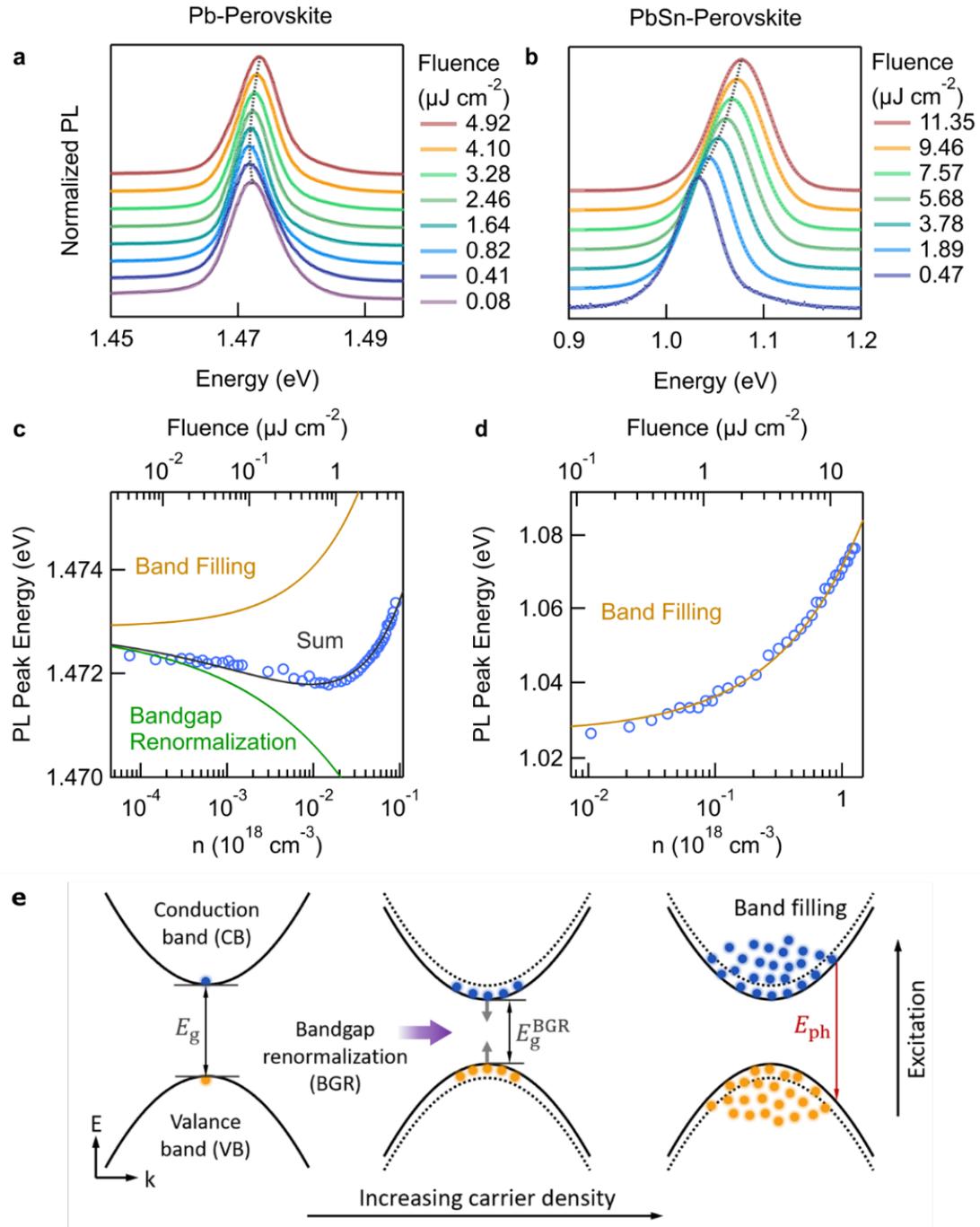

**Figure 5.** Band filling and bandgap renormalization effects. **a** Normalized PL spectra of the Pb-perovskite with excitation fluence increasing from 3 nJ cm$^{-2}$ to 5 μJ cm$^{-2}$ at 10 K, showing a small redshift at low fluence and then a blueshift at high fluence. **b** Fluence-dependent PL spectra of the PbSn-perovskite with excitation fluence from 129 nJ cm$^{-2}$ to 15 μJ cm$^{-2}$, showing a large and monotonic blueshift. Shifts of the PL peak energy as a function of the excitation fluence for the Pb- (**c**) and PbSn- (**d**) perovskites, respectively. The corresponding carrier densities (*n*) are indicated in the bottom axis. Blue circles represent the experimental data. The brown line shows the band filling effect ($E_g^0 + \Delta E_g^{BM}$), the green line shows the bandgap renormalization effect ($E_g^0 + \Delta E_g^{BGR}$), and the black line is a sum of both ($E_g^0 + \Delta E_g^{BM} + \Delta E_g^{BGR}$), where $E_g^0$ is the intrinsic bandgap that does not depend on *n*. **e** Schematic of the bandgap renormalization and band filling effects.

The band filling effect, also called the Burstein-Moss shift[38,39], has been widely observed in perovskites by transient absorption[37] and PL spectroscopy[15]. As schematically shown in Figure 5e, a relatively low density of photoinduced carriers can fill the bands to an appreciable depth and thus can widen the effective bandgap, inducing optical transitions at energies higher than the fundamental bandgap. Because of the similar effective mass of electrons and holes in metal trihalide perovskites[63], both carriers contribute to the state filling. The magnitude of the bandgap widening due to the band filling effect can be modelled by the Burstein-Moss expression[64] $\Delta E_g^{BM} = \hbar^2(3\pi^2 n)^{2/3}/2\mu$, where $\hbar$ is the reduced Plank constant and $\mu$ is the reduced effective mass of excitons.

Although the band-filling effect in the perovskites was detected first by the shift in the absorption edge upon photoinjection[37], the large PL blue shift has been observed in FASnI$_3$ and has been attributed to the slow hot carrier cooling[15]. The possible phonon bottleneck effect[62] makes the PL emission energy mostly contributed by the recombination of the electrons and holes near their quasi Fermi levels, whose difference is therefore a good representation of the optical gap.

Besides the intuitive filling process, optical excitations can generate a dense electron-hole plasma where many-body effects take place. Carriers with the same charge repel each other by the Coulomb interaction, and the Pauli exclusion principle further separates carriers with the same spin. The correlation and exchange interaction between the carriers cause screening in the electron-hole system and renormalizes their self-energies, which narrows the bandgap[40,41] (Figure 5e).

The BGR effect has been investigated in-depth in heavily doped semiconductor systems. A phenomenological expression[65] describes the BGR energy shift as $\Delta E_g^{\text{BGR}} = -\gamma n^{1/3}$, where $\gamma$ is the BGR coefficient related to the static dielectric constant of the material. The total change in the bandgap can be defined as a combination of the two contributions $\Delta E_g = \Delta E_g^{\text{BM}} + \Delta E_g^{\text{BGR}}$, yielding a redshift of the optical bandgap at low fluences and then a blueshift at high fluences.

Figure 5c shows an excellent description of the Pb-perovskite data over the entire range of excitation intensities when both the band-filling and BGR effects are considered. From the fit, we can estimate $\mu = 0.146(6)m_0$ for the Pb-perovskite, similar to that of $0.1m_0$ determined by high-field magnetoabsorption measurements[47], where $m_0$ is the free electron mass. We also obtain a BGR coefficient of $\gamma = 1.05(5) \times 10^{-8}$ eV cm, which is comparable to the values in typical inorganic semiconductors, such as $4.27 \times 10^{-8}$ eV cm in Si-doped GaN (Ref.[66]), $2.25 \times 10^{-8}$ eV cm in n-type InP (Ref.[67]), and $1.2 \times 10^{-8}$ eV cm in lead-iodide perovskites extracted by transient absorption spectroscopy[62]. Previous fluence-dependent studies[15,37] for the Pb-perovskites injected photocarriers higher than $n\sim10^{16}$ cm$^{-3}$, obtaining a much higher exciton effective mass of $0.33m_0$; our work reveals that it is critical to cover very low excitation densities to successfully account for the BGR effect in the effective mass calculation (a comparison with one excluding the BGR effect is demonstrated in Supporting Note 9).

In the PbSn-perovskite, the large dark noise of the InGaAs detector limits our measurement range of photoexcitation densities to $n\sim10^{16}$ cm$^{-3}$ and above. This restricts the PL detection for the PbSn-perovskite to the fluence range where band filling is dominant. In Figure 5d, the PbSn-perovskite data can be fit perfectly with the Burstein-Moss expression, confirming the dominance of band filling over the entire fluence range. The obtained $\mu = 0.081(1)m_0$ agrees well with the previous report of $0.12m_0$ at 24 K in FASnI$_3$ obtained from power-dependent PL spectra[15] and the calculation result of $0.075m_0$ in mixed-PbSn perovskites[63].

## 3. Conclusion

In summary, we report on the temperature and photoexcited carrier density dependence of the bandgap in two phase-stabilized perovskite thin films $MA_{0.3}FA_{0.7}PbI_3$ and $MA_{0.3}FA_{0.7}Pb_{0.5}Sn_{0.5}I_3$. In stark contrast to previous studies, we have revealed that the temperature-dependent bandgap shifts in our samples cannot be explained by the thermal lattice expansion and structural phase transitions, due to their small lattice expansion and stabilized cubic structure. Instead, using first-principles simulations, we have proposed that the unusual bandgap shift with temperature originates from static local distortions at low temperatures and their competition with dynamic lattice vibrations. By measuring fluence-dependent PL down to a very low excitation density ($10^{14}$ $cm^{-3}$), we have demonstrated the BGR effect for the first time in addition to the band-filling effect, and have shown that accounting for the BGR effect leads to a more accurate estimation of the exciton effective mass. Our findings clarify and provide a deeper understanding of the optical bandgap shifts with temperature and excitation intensity in perovskite thin films, and are instructive for future applications and designs of perovskite-based optoelectronic devices.

## 4. Experimental Section

**Perovskite Thin Film Preparation**

The samples in this study are solution-processed $MA_{0.3}FA_{0.7}PbI_3$ and $MA_{0.3}FA_{0.7}Pb_{0.5}Sn_{0.5}I_3$ perovskite thin films (hereafter referred to as Pb- and PbSn-perovskite). The PbSn-perovskite was prepared using a tin-reduced precursor solution strategy to effectively prevent the oxidation of $Sn^{2+}$ to $Sn^{4+}$. This synthesis strategy, which is described in Ref.[42], significantly reduces both the trap density and the hole density by a factor of two compared to a control sample. The thicknesses of the Pb- and the PbSn-perovskite samples are 1.0 μm and 1.2 μm, respectively (see Figure S10, Supporting Information).

**X-Ray Diffraction Spectroscopy**

Crystal structure and lattice parameters were determined by X-ray diffraction using a Bruker D8 Advanced diffractometer with Cu Kα radiation (wavelength: 1.54 Å). Samples were transported to a liquid nitrogen-cooled stage under vacuum and operated in a temperature range of 90 K to 340 K.

**Temperature-Dependent Absorption and Photoluminescence**

The thin films were mounted on a cold finger of a closed-loop helium cryostat under vacuum to vary the temperature from 10 K to 340 K. For absorption measurements, a stabilized quartz-tungsten lamp was used as the white light source. The beam was spatially limited by an iris, collimated, and then focused on the sample with a 50 mm focal length lens. The transmitted light was collected by a multi-mode fiber coupled to the entrance slit of a Czerny-Turner spectrometer. The spectra were recorded by a thermoelectrically cooled EMCCD (detection range: 250-1050 nm) for the Pb-perovskite and an InGaAs single-point detector (800-2500 nm) for the PbSn-perovskite. For temperature-dependent PL measurements, two continuous-wave diode lasers (633 nm and 785 nm for the Pb- and the PbSn-perovskites, respectively) were used as the excitation sources. The setups were similar to our previous study[68].

**Fluence-Dependent Photoluminescence**

The fluence-dependent PL was measured with an 80-MHz mode-locked Ti:Sapphire laser for accurate calculations of the photoinjected carrier density. The beam passed through an acoustic-optical pulse selector with a 20:1 division ratio to reduce the thermal effect and achieve a full recovery from the excitation. The Pb- and PbSn-perovskites were excited at the wavelength of 720 nm and 785 nm, respectively. The maximum power on the PbSn-perovskite is larger than that on the Pb-perovskite because the peak power output from the Ti:Sapphire laser is located at ~805 nm. The laser power incident on the sample was adjusted by a set of variable neutral density filters. Both the excitation light illumination and the PL collection were on the thin-film side rather than the substrate side to avoid self-absorption effect. Notably, the EMCCD camera provides exceptional sensitivity and noise performance (~single electron) compared to the InGaAs detector (~$10^3$ electrons). Therefore, only the PL of the Pb-perovskite can be detected with a laser fluence down to 3 nJ cm$^{-2}$.

**In Situ High-Pressure Photoluminescence**

High-pressure experiments were carried out using a symmetric diamond anvil cell apparatus. The samples were scraped off from the substrates and loaded into a 250 μm-diameter chamber, which was constructed from a T301 steel gasket and pre-indented into a thickness of 40 μm. Mineral oil was used as the pressure-transmitting medium. Several ruby balls with diameters of approximately 30 μm were placed in the sample chamber. The pressure calibration was determined by the ruby fluorescence method. A diode laser with an excitation wavelength of 633 nm was used for all luminescence experiments. The beam was focused onto the sample using a 20× long working distance objective.

**Computation Methods**

DFT calculations were performed with the Vienna *ab initio* simulation package[69] using the projector-augmented wave method together with the plane-wave basis with an energy cutoff of 430 eV. The Perdew-Burke-Ernzerhof exchange-correlation functional[70] was applied. The materials studied by experiments have mixed organic components, which are difficult to handle in DFT calculations. Since the A-site molecule has minor influence on electronic states near the Fermi level, we simplified the system from the mixed A-site to pure MA and include the influence of FA by adding structural constraints. That is, the structure was artificially kept in cubic phase during relaxation and the experimental lattice constants from our XRD measurements were employed, which were fixed to be 6.375 Å for $MAPbI_3$ and 6.410 Å for $MAPb_{0.5}Sn_{0.5}I_3$. The MA molecule was aligned along the [100] direction, as recommended by previous work[34]. $MAPb_{0.5}Sn_{0.5}I_3$ was simulated by doubling the unit cell along the [100] direction and replacing one Pb with Sn. Structural relaxation was performed with fixed lattice constants, using a force convergence criterion of $10^{-2}$ eV Å$^{-1}$. The k-points grids were set as 4 × 4 × 4 for $MAPbI_3$ and 2 × 2 × 2 for $MAPb_{0.5}Sn_{0.5}I_3$ for structural relaxation, which were increased to 8 × 8 × 8 for self-consistent calculations. A van der Waals correction was introduced by the DFT-D3 method[71] with the Becke-Jonson damping[72].

**Acknowledgements**

Samples were prepared at Nanjing University. All optical measurements and first-principles calculations were performed at Tsinghua University. L.Y. acknowledges the support from the National Key R&D Program of China (Grant Nos. 2020YFA0308800 and 2021YFA1400100) and the National Natural Science Foundation of China (Grant Nos. 12074212). Y.X. was supported by the National Key R&D Program of China (2018YFA0307100 and 2018YFA0305603) and the National Natural Science Foundation of China (12025405 and 11874035). The work of H.T. was supported by the National Natural Science Foundation of China (Grant Nos. 61974063 and U21A2076). H.Z. was also supported by funds from the University of Toronto.

## Competing interests

The authors declare no competing interests.